\documentclass[journal=cmatex,manuscript=article]{achemso}
\usepackage{chemformula} 
\usepackage[T1]{fontenc} 

\author{Steven R. Spurgeon}
\affiliation{Energy and Environment Directorate, Pacific Northwest National Laboratory, Richland, Washington 99352}
\email{steven.spurgeon@pnnl.gov}

\author{Tiffany C. Kaspar}
\affiliation{Physical and Computational Sciences Directorate, Pacific Northwest National Laboratory, Richland, Washington 99352}

\author{Vaithiyalingam Shutthanandan}
\affiliation{Environmental Molecular Sciences Laboratory, Pacific Northwest National Laboratory, Richland, Washington 99352}

\author{Jonathan Gigax}
\affiliation{Department of Nuclear Engineering, Texas A\&M University, College Station, Texas 77843}

\author{Lin Shao}
\affiliation{Department of Nuclear Engineering, Texas A\&M University, College Station, Texas 77843}

\author{Michel Sassi}
\affiliation{Physical and Computational Sciences Directorate, Pacific Northwest National Laboratory, Richland, Washington 99352}

\title{Asymmetric Lattice Disorder Induced at Oxide Interfaces}

\begin{document}

\begin{tocentry}

\includegraphics[width=\textwidth]{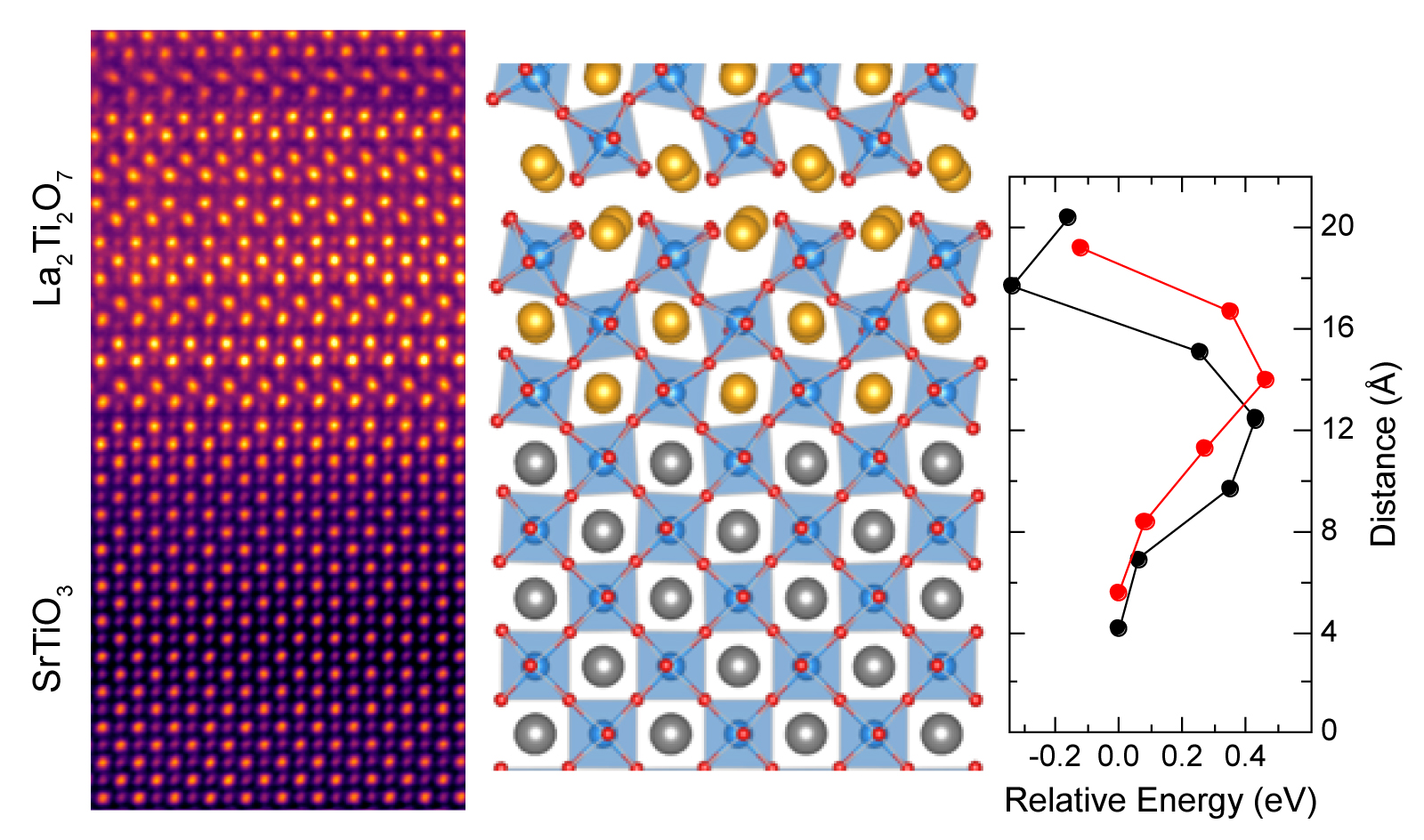}
Correlative atomic-scale imaging and theory calculations reveal the presence of asymmetric oxygen vacancy formation during disordering of complex oxide interfaces.

\end{tocentry}

\begin{abstract}

Control of order-disorder phase transitions is a fundamental materials science challenge, underpinning the development of energy storage technologies such as solid oxide fuel cells and batteries, ultra-high temperature ceramics, and durable nuclear waste forms. At present, the development of promising complex oxides for these applications is hindered by a poor understanding of how interfaces affect lattice disordering processes and defect transport. Here we explore the evolution of local disorder in ion-irradiated La$_2$Ti$_2$O$_7$ / SrTiO$_3$ thin film heterostructures using a combination of high-resolution scanning transmission electron microscopy (STEM), position-averaged convergent beam electron diffraction (PACBED), electron energy loss spectroscopy (STEM-EELS), and \textit{ab initio} theory calculations. We observe highly non-uniform lattice disordering driven by asymmetric oxygen vacancy formation across the interface. Our calculations indicate that this asymmetry results from differences in the polyhedral connectivity and vacancy formation energies of the two interface components, suggesting ways to manipulate lattice disorder in functional oxide heterostructures.

\end{abstract}

\section{Introduction}

Thin film oxide heterostructures have emerged as key enabling technologies for transformative advances in electronics, quantum computing, and energy production. These systems possess a range of tunable functionalities, such as conductivity and magnetism, that result from non-equilibrium growth conditions, are distinct from the bulk, and are heavily influenced by lattice disorder.\cite{Martin2010} Order-disorder phase transitions are particularly important in  $A_2B_2$O$_7$ compounds (where $A$ and $B$ are transition metal cations), which have been studied for use in solid oxide fuel cells (SOFCs),\cite{Ignatiev2016, Jacobson2012, Santiso2011} advanced ferroelectric sensors,\cite{Kaspar2018,Yan2009} and nuclear waste forms.\cite{Sickafus2000} These materials represent model systems in which to examine the fundamental relationship between the introduction of point defects and lattice disorder, which together govern behavior.

In the bulk, $A_2B_2$O$_7$ compounds can exhibit a range of structures from defected fluorite ($Fd\bar{3}m$) to cubic pyrochlore ($Fd\bar{3}m$) and monoclinic perovskite-like layered structure ($P2_1$), as the cation radius ratio ($r_A/r_B$) increases.\cite{Subramanian1983} During thin film growth, additional kinetic and substrate constraints can further change the energy landscape of the system and the resultant phase distribution.\cite{Kaspar2018, Kaspar2017} This structural flexibility allows the pyrochlores to preserve their overall crystallinity during disordering of the cation sublattice by external stimuli, such as electric fields and energetic ion bombardment, making them promising candidates for SOFCs and nuclear waste forms.\cite{Shamblin2016, Park2015, Lang2010} Recently, Kreller \textit{et al.}\cite{Kreller2019} have shown that ion irradiation may be used to controllably manipulate anti-site defect populations and resulting ion conduction at a fundamental level. Building upon this approach, ion irradiation applied to thin film heterostructures represents a promising route to examine how interface geometry and configuration mediate the evolution of local order. Furthermore, characterization using aberration-corrected scanning transmission electron microscopy can provide direct insight into defect formation that can be linked to theory models.

Decades of work on metals have shown that grain boundaries and other interfaces can significantly influence point defect transport and annihilation,\cite{Odette2008} motivating the use of engineered nanostructured materials to mediate disordering processes.\cite{Zhang2018a, Beyerlein2015, Beyerlein2013} While irradiation-induced structural disorder has been extensively studied in bulk pyrochlores,\cite{Lang2010, Lian2003, Lian2003a, Sickafus2000} less attention has been paid to the behavior of irradiated oxide thin film interfaces. Previous studies of epitaxial TiO$_2$,\cite{Zhuo2012, Zhuo2011} CeO$_2$,\cite{Aguiar2014, Dholabhai2014} and BaTiO$_3$\cite{Bi2013} films on SrTiO$_3$ have shown evidence for the presence of denuded zones in some interface configurations and not in others; these variations have been attributed to differences in defect kinetics and energetics of the interface and its component elements. The interface charge state is also thought to play a role in determining the stability of SrTiO$_3$ / MgO interfaces.\cite{Aguiar2014a} In addition, vacancy transport is an important factor in both disordering processes and ion conduction.\cite{Uberuaga2019, Uberuaga2018, Dholabhai2017, PedroUberuaga2013} Recently, we showed that epitaxial La$_2$Zr$_2$O$_7$ thin films with the defected fluorite structure amorphized under heavy ion irradiation starting in the surface region, then at the film/substrate interface, and that the central portion of the film exhibited higher radiation tolerance.\cite{Kaspar2017} At present it is not clear what factors dictate the response of oxide heterostructures, limiting our ability to direct their behavior. Understanding how the lattice geometry of oxide phases impacts defect formation and transport at interfaces may provide a pathway to control functionality in these materials.

Here we examine ion-irradiation-induced disordering of La$_2$Ti$_2$O$_7$ (LTO) / SrTiO$_3$ (STO) interfaces using a combination of aberration-corrected scanning transmission electron microscopy (STEM), position-averaged convergent beam electron diffraction (PACBED), and electron energy loss spectroscopy (STEM-EELS), supported by \textit{ab initio} simulations. We observe a highly non-uniform disordering process in the heterostructure, with amorphization of the film and substrate, but with retained interface crystallinity and local order at the highest doses examined. Electronic fine structure measurements reveal the presence of significant oxygen vacancy populations, whose energetic stability and spatial distribution are evaluated by \textit{ab initio} calculations. Our results emphasize the role of the polyhedral network in determining the energetics of oxygen vacancy formation and in mediating order-disorder phase transitions at interfaces. Collectively, these findings offer ways to tune such transitions through interface engineering, with broad implications for device properties and performance in extreme environments.

\section{Results and Discussion}

Several nominally 50 nm-thick epitaxial La$_2$Ti$_2$O$_7$ thin films were deposited on SrTiO$_3$ (110) substrates using pulsed laser deposition (PLD) and annealed in air,\cite{Kaspar2018} as described in the Methods. Ion irradiations were performed using 1 MeV Zr$^+$ ions applied 7$^{\circ}$ off-normal to three quadrants of the same $1 \times 1$ cm$^2$ sample, yielding four conditions spanning 0, 1, 2, and 4 dpa. Cross-sectional STEM samples were extracted from each region and imaged in the high-angle annular dark field (STEM-HAADF) mode, in which contrast is proportional to atomic number ($Z^{\sim1.7}$). This mode is sensitive to local atomic structure and can be used to directly resolve interface configurations as an input for modeling. Rutherford backscattering spectrometry (RBS) measurements using 2 MeV He$^{+}$ were also performed before and after irradiation to assess area-averaged changes in crystallinity.

\begin{figure}
\includegraphics[width=0.7\textwidth]{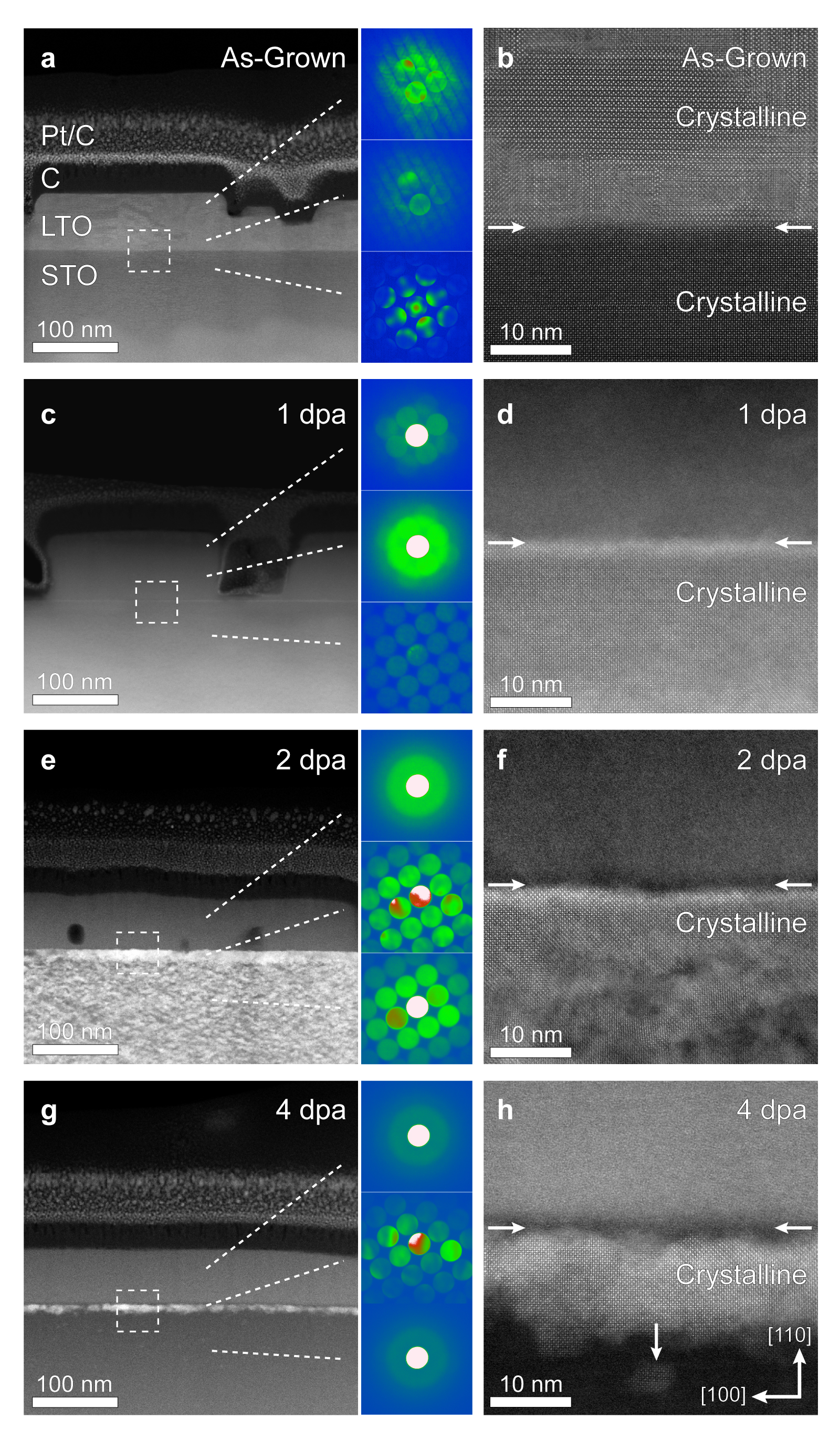}
\caption{Visualizing structural disorder. Representative low-magnification and high-magnification STEM-HAADF images taken from the dashed square, inset with PACBED from the marked regions, for as-grown (a--b), 1 (c--d), 2 (e--f), and 4 (g--h) dpa samples. The horizontal arrows mark the position of the interface, while the vertical arrow in (h) indicates a small crystalline patch deeper into the substrate. \label{haadf}}
\end{figure}

As shown in Figure \ref{haadf}.a--b, the heterostructure before irradiation is fully crystalline and epitaxial, consisting of large regions of the well-ordered La$_2$Ti$_2$O$_7$ monoclinic perovskite-like layered structure (vertical or horizontal stripe contrast that corresponds to the additional (110) oxygen planes that occur between every four perovskite-like units of LTO\cite{Kaspar2018}), as well as some smaller regions of cubic perovskite LaTiO$_3$. The interface appears relatively sharp, with two predominant configurations that will be described later. To gain further insight into local crystallinity as a function of irradiation dose, we performed position-averaged convergent-beam electron diffraction, as shown in the center panels in Figure \ref{haadf}. While conventional selected-area electron diffraction (SAED) has been used to examine disorder in irradiated pyrochlores,\cite{Qing2018,Lian2003a,Lian2002,Lian2003} this approach is only suited to probing larger grains of material ($>0.1$ \textmu m) and is difficult to selectively apply to different parts of a nanoscale-sized thin film. PACBED enables direct examination of crystallinity and structural disorder at the nanoscale and can be related to other chemical information from STEM.\cite{Janish2019, Ophus2019} PACBED patterns from the unirradiated film and substrate (Figure \ref{haadf}.a) confirm the presence of full crystalline ordering throughout, indicated by overlapping circular diffraction disks in the patterns.\cite{Ophus2017b,LeBeau2010} While a relaxed probe convergence angle and small condenser aperture were used, some disk overlap could not be avoided for the crystal structure of the film. The stripes in the patterns result from the superlattice periodicity of oxygen defect planes in the structure, which are shown in detail in the high-magnification image in Figure \ref{haadf}.b. After irradiation with 1 MeV Zr$^+$ to a dose of 1 dpa, the diffraction disks in the PACBED pattern of the LTO are superimposed on a broad ring of more uniform intensity, indicating the coexistence of both ordered and disordered regions in the LTO film (Figure \ref{haadf}.c). In contrast, the STO side of the interface remains largely crystalline, as confirmed by both the diffraction pattern and high-magnification images (Figure \ref{haadf}.d). At a dose of 2 dpa, the LTO PACBED pattern shows a diffuse ring pattern with no residual diffraction disks (Figure \ref{haadf}.e). These results are in agreement with RBS measurements, shown in Figure S1, which indicate that some crystallinity in the LTO is retained up to 1 dpa, after which the film becomes largely disordered. Both low- and high-magnification images show that the bulk of the substrate is increasingly disordered. In contrast, an obvious 10--15 nm-thick crystalline band is retained on the STO side of the interface (Figure \ref{haadf}.f). Finally, at the highest dose of 4 dpa, PACBED patterns collected from the LTO near the interface and STO far from the interface are increasingly diffuse and uniform (Figure \ref{haadf}.g), reflecting an amorphous structure. However, the strongly diffracting crystalline band of STO at the interface is even more pronounced in this case, as are some small crystalline patches deeper into the substrate side (marked by the second arrow in Figure \ref{haadf}.h). It is important to emphasize that these differences are not the result of a non-uniform ion implantation profile, since ion channeling simulations\cite{Kaspar2017} using the SRIM code\cite{Ziegler2010} confirm a fairly flat distribution of damage within the first 50 nm of the sample for the chosen ion species and energy. Rather, these results show that the character and distribution of defects formed is strongly influenced by the presence of the interface.

\begin{figure}
\includegraphics[width=\textwidth]{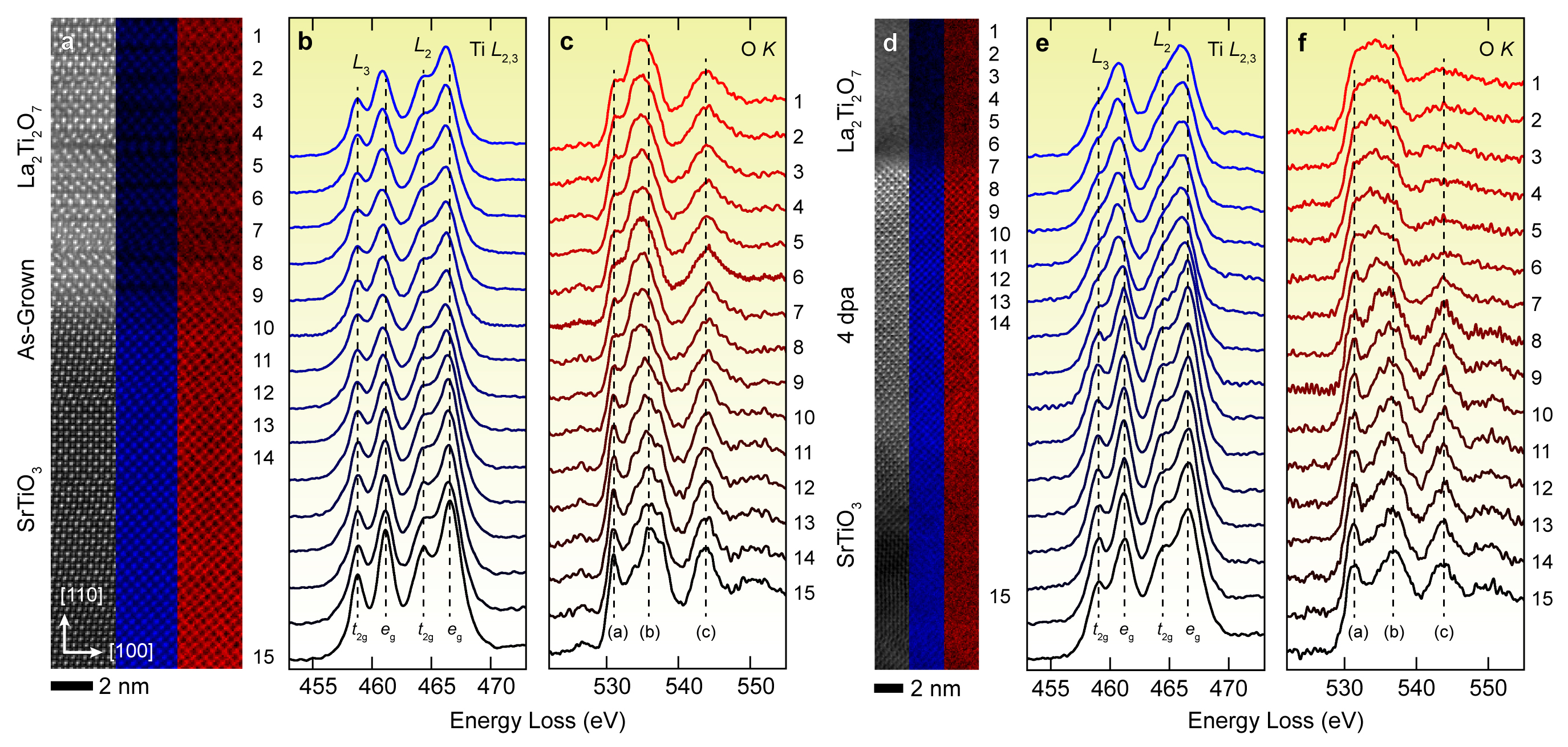}
\caption{Changes in local defect environment. STEM-HAADF images and STEM-EELS maps of the Ti $L_{2,3}$ (blue) and O $K$ (red) edges for the unirradiated (a) and 4 dpa (d) samples. Corresponding spectra collected from the marked regions of the interface for the unirradiated (b--c) and 4 dpa (e--f) samples. The maps have been processed using principal component analysis (PCA) to improve signal-to-noise, but the individual spectra have not been denoised in any way, as described in the methods.\label{eels}}
\end{figure}

To further explore the evolution of chemical disorder at the film-substrate interface, we have performed atomic-scale electron energy loss spectroscopy measurements. STEM-EELS has long been used to explore changes in the bonding, chemistry, and defects at oxide interfaces.\cite{Gazquez2016, Tan2012, Varela2006a} This technique can be quantitatively interpreted through \textit{ab initio} calculations, yielding powerful predictive insight into the behavior of complex oxides.\cite{MacLaren2014, Keast2001} In particular, we focus on the Ti $L_{2,3}$ and O $K$ edge fine structure in the unirradiated and 4 dpa samples. These edges result from spin-orbit split Ti $2p_{3/2}$, $2p_{1/2} \rightarrow 3d$ and O $1s \rightarrow 2p$ state transitions, respectively, and encode rich information about the cation coordination and defect chemistry in the LTO / STO system.\cite{Mosk2004} Figure \ref{eels}.a--c show maps and spatially-resolved fine structure measurements collected from LTO and STO near the interface, as well as the bulk of the STO. The distribution of features in the Ti $L_{2,3}$ edge, particularly the crystal field splitting of the $t_{2g}$ and $e_g$ peaks for both the $L_3$ (460 eV) and $L_2$ (465 eV) edges, change upon moving from STO to LTO due to the different environment of substrate and film. The overall shape and reduced peak-to-valley features of the edge are in reasonable agreement with prior measurements on La$_2$Ti$_2$O$_7$ and STO.\cite{Ohtomo2002} Changes are also present at the O $K$ edge, which displays an decreased pre-to-main peak ratio, observed by comparing feature (a) ($\sim 531$ eV) and (b) ($\sim536$ eV) in Figure \ref{eels}.c, as well as a shift of the main peak (b) and broadening of feature (c) at 544 eV relative to STO. The pre-peak feature (a) has been shown to have a strong $B$-site cation $3d$ band contribution in similar oxides, while the main peak (b) results from hybridization and is influenced by the occupancy of the $A$-site cation $d$ bands, making these features useful measures of local defect populations.\cite{Varela2009} Oxygen vacancies are known to have a large effect on the fine structure, acting both to reduce the crystal field splitting of the Ti $L_{2,3}$ edge, as well as damping out the oscillations in the O $K$ edge.\cite{Mosk2004} Measurements of the 4 dpa sample, shown in Figure \ref{eels}.d-f, indicate extensive local structural disorder and oxygen vacancy formation. Focusing first on the LTO side, we note the strong blurring of the $t_{2g}$ and $e_g$ features in the Ti edge, reflecting the highly disordered nature of the LTO cation network. This behavior is mirrored by the O $K$ edge, in which the there is no longer any separation between pre- (a) and main (b) peaks and the third peak (c) is almost entirely dampened, resulting from the loss of long-range order. The LTO is fully amorphous down to the interface, with minimal La / Sr intermixing after irradiation. While a clear crystalline band is retained on the STO side of the interface, its fine structure is completely changed relative to the starting material. There is a reduction in the $t_{2g} : e_g$ peak ratio for both the $L_3$ and $L_2$ edges, which indicates a reduction in Ti valence from 4+ to 3+ due to the presence of increased oxygen vacancies.\cite{Mosk2004} Similarly, there is a pronounced damping and general broadening of all the features in the O $K$ edge due to increasing disorder in the oxygen network. These results show that, although the overall crystallinity of STO is retained, there are nonetheless substantial changes in electronic structure in this region.

\begin{figure}
\includegraphics[width=\textwidth]{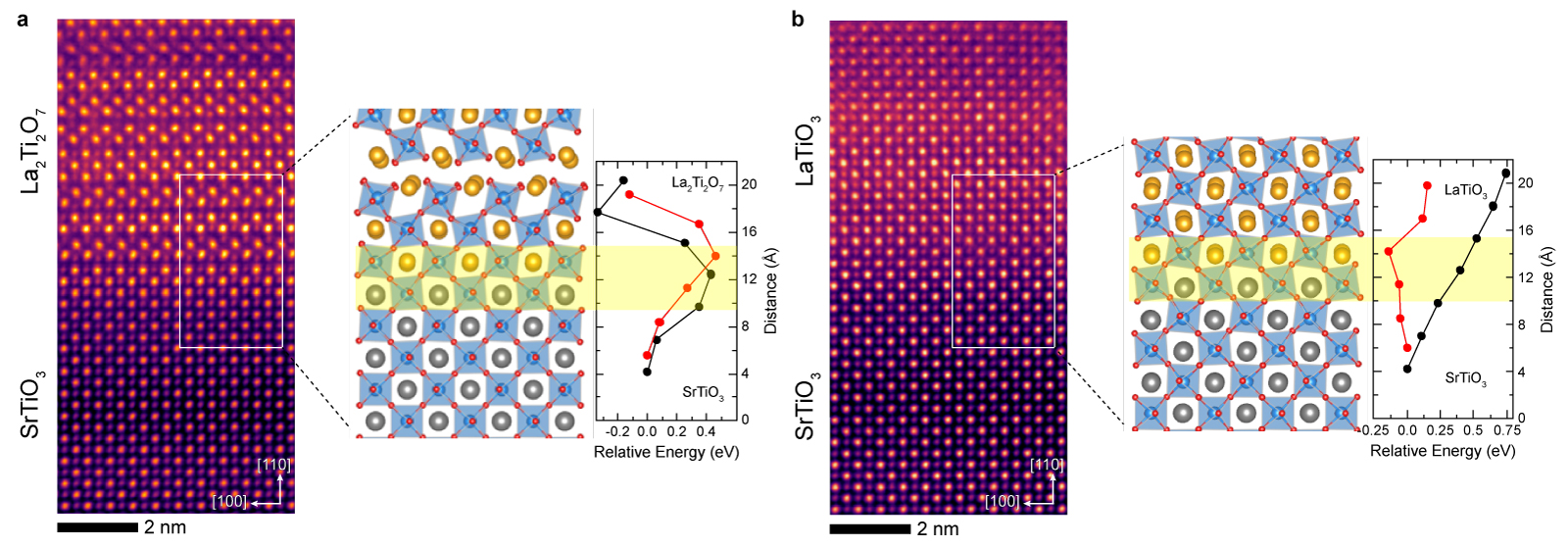}
\caption{Asymmetry in oxygen vacancy formation. STEM-HAADF images of two interface configurations observed in unirradiated LTO / STO (110) and corresponding \textit{ab initio} theory calculations for the energy of oxygen vacancy formation as a function of position in the majority La$_2$Ti$_2$O$_7$ / STO (a) and minority LaTiO$_3$ / STO (b) interface configurations. \label{dft}}
\end{figure}

These structural and chemical results illustrate the complex, non-uniform evolution of nanoscale disorder and the unique response of the LTO / STO interface. We have examined the interface configuration of the starting heterostructure in detail to explore possible defects underpinning the observed disordering behavior. As shown in Figure \ref{dft}, we observe two main types of LTO / STO interface: a predominant perovskite-like layered La$_2$Ti$_2$O$_7$ / STO configuration and some small regions of a cubic perovskite LaTiO$_3$ / STO configuration. High-magnification STEM-HAADF images provide direct atomic-scale interface configurations, which we use as the basis for \textit{ab initio} calculations to explore the energetics of oxygen vacancy ($V_O$) formation. Oxygen vacancy defect formation is common in ion-irradiated oxides,\cite{Sickafus2000} and their transport in perovskites\cite{Uberuaga2015} and pyrochlores\cite{Uberuaga2019} is an area of ongoing study. In the case of the more abundant La$_2$Ti$_2$O$_7$ / STO interface, the relative energy diagram shown in Figure \ref{dft}.a indicates that it is more favorable to form an oxygen vacancy in the LTO film, followed by STO substrate, and then at the interface. This trend suggests that there is a thermodynamic driving force for accumulation of oxygen vacancies within the additional oxygen planes of the LTO film. In these interlayers there is a discontinuity in the connectivity of the Ti polyhedral network, which has been shown to affect the migration and final position of vacancies in monoclinic perovskite-like layered\cite{Yang2016} and cubic pyrochlore\cite{Uberuaga2019, Sassi2019} structures. While multiple factors play a role in the amorphization of ion-irradiated materials and subsequent mass transport, the relative trend in the energetically favorable position of oxygen vacancy formation is consistent with the amorphization sequence observed experimentally in Figure \ref{haadf}: the LTO film amorphizes first, followed by STO, while the STO at the interface remains mostly crystalline, albeit still containing defects. These calculations are also in line with the large oxygen vacancy populations observed in STEM-EELS. To confirm that this amorphization behavior is due (at least in part) to the ability of LTO to absorb oxygen vacancies, we compare these results against the trend calculated for the minority LaTiO$_3$ / STO interface, shown in Figure \ref{dft}.b. We note that because most of the LTO has amorphized at 1 dpa, we cannot independently track the behavior of these smaller regions relative to the La$_2$Ti$_2$O$_7$ / STO regions in the STEM. However, for this configuration we find that the energy cost associated with defect formation at the interface and STO substrate side are lower than for the LaTiO$_3$ film, indicating that the film should amorphize last. In this configuration there is no discontinuity in the polyhedral network that would help to stabilize oxygen vacancies. Rather, it is more likely that the charge state of the interface and the response of the individual components govern the tendency to form vacancies. Aguiar \textit{et al.}\cite{Aguiar2014a} have previously examined the behavior of epitaxial STO (110) and (100) on MgO (100), finding that the former interface tends to amorphize more readily. They attributed this behavior to the charged nature of the (110) planes in STO, which may act to destabilize ions within the planes under irradiation. However, the STO orientation is the same for the two interface configurations we have examined, underscoring the effect of crystal structure on vacancy formation. Taken together, these results emphasize the important role of the lattice polyhedral network and suggest that it may be used as a means to direct the formation of oxygen vacancies at interfaces. 

\section{Conclusions}

Our results show that order-disorder phase transitions in complex oxides can be strongly influenced by the presence and configuration of interfaces. We find that monoclinic perovskite-like layered perovskite La$_2$Ti$_2$O$_7$ structures with varied polyhedral connectivity can act as sinks for oxygen vacancy formation when compared to the uniform, cubic SrTiO$_3$ perovskite. At the doses studied, the sink behavior of LTO strongly influences the amorphization behavior of the STO at the interface; this region retains its crystallinity when both the LTO and deeper STO have amorphized. These results suggest that it may be possible to tune and control lattice disorder through different combinations of layered structures, or perhaps engineering of oxygen-deficient structures such as Brownmillerite phases.\cite{Parsons2009} Careful consideration should be given to the integration of components for SOFCs and other applications, recognizing that substrate-induced deviations in phase can affect the stability and transport of oxygen vacancies. Importantly, our results emphasize that relative defect formation tendencies in interface components is likely a major driver for the observed asymmetric defect behavior.

\section{Experimental Section}

\subsection{Synthesis and Irradiation}

Nominally 50 nm-thick epitaxial La$_2$Ti$_2$O$_7$ (LTO) thin films were deposited by pulsed laser deposition at 950 $^{\circ}$C and 0.5 mTorr pO$_2$ on SrTiO$_3$ (110) substrates, as described previously.\cite{Kaspar2018} Post-deposition annealing at 1100$^{\circ}$ in air for 4 h was necessary to obtain the perovskite-like layered structure (PLS) crystal phase. Annealing also resulted in an increase in surface roughness and, in some cases, voids within the film, due to bulk mass transfer during the annealing process. The films were irradiated with 1 MeV Zr$^{+}$ produced by a 1.7 MV tandem accelerator at the Texas A\&M University Ion Beam Lab. Zr$^+$ ions were implanted at 7$^{\circ}$ off normal and room temperature to the desired dose (1, 2, and 4 dpa, corresponding to ion fluences of $7.96 \times 10^{13} - 3.19 \times 10^{14}$ ions cm$^{-2}$, respectively); each dose was implanted in a separate quadrant of the same film. Fluence and dpa were calculated as described previously for La$_2$Zr$_2$O$_7$,\cite{Kaspar2017} with no adjustments for LTO. Films were analyzed before and after irradiation by Rutherford backscattering spectrometry (2 MeV He$^+$) in the random and channeling (RBS/c) directions.

\subsection{Scanning Transmission Electron Microscopy}

STEM samples were prepared using a FEI Helios NanoLab DualBeam Focused Ion Beam (FIB) microscope and a standard lift out procedure. STEM-HAADF images and STEM-EELS fine structure maps were collected on a probe-corrected JEOL ARM-200CF microscope operating at 200 kV accelerating voltage, with a probe semi-convergence angle of 27.5 mrad, a collection angle of 82--186 mrad, and a EELS inner collection angle of 42.9 mrad. The images in Figure \ref{dft} were acquired from a series of 10 images aligned using the SmartAlign program\cite{Jones2015} and filtered using an average background subtraction filter to improve signal-to-noise. EELS fine structure maps were collected using a 1~\AA~probe size with a $\sim130$ pA probe current and a 0.25 eV ch$^{-1}$ dispersion, yielding an effective energy resolution of $\sim0.75$~eV. The O $K$ and Ti $L_{2,3}$ edge spectra were corrected for energy drift using the zero-loss peak, then treated with a power low background subtraction fit to a 30 and 25 eV window prior to the edges, respectively, and processed using a Fourier-Ratio deconvolution. The maps shown in Figure \ref{eels} were processed using principal component analysis to improve signal-to-noise, but the extracted spectra are not denoised in any way. Energy-filtered PACBED patterns were collected on a JEOL GrandARM-300F microscope operating at 300 kV with $\alpha = 1$, yielding a relaxed convergence semi-angle of $\sim4$ mrad using the spectrometer camera with an 8 cm camera length and a 10--20 eV spectrometer entrance slit to increase contrast. Patterns were averaged over 5--10 nm$^2$ regions from the top, middle, and bottom of the sample.

\subsection{Theory Calculations}

Computational models of the LTO / STO interfaces used a periodic symmetric slab such that LTO was sandwiched on both sides of a 4.5 unit cell STO slab (i.e. LTO / STO / LTO). Convergence of the relative energy for oxygen vacancy formation was tested against the number of LTO unit cells used to model the interface. The optimal number of unit cells to model the La$_2$Ti$_2$O$_7$ / STO interface was found to be 2, while 4 unit cells were required for the LaTiO$_3$ / STO interface. The lattice parameters of the simulation box used $a \sqrt{2} \times a \sqrt{2}$ as lateral cell and a vacuum space of 40~\AA \, in the out-of-plane direction to avoid interaction between periodic images. The lattice parameters were fixed and used the experimental lattice parameter for STO ($a=3.905$~\AA). In order to avoid artificial dipole effects, oxygen vacancies were symmetrically created in LTO on both sides of the STO slab. The relative formation energy of oxygen vacancy was calculated using the density functional theory framework, as implemented in the VASP code.\cite{Kresse1996} All the simulations used the PBEsol density functional\cite{Perdew2008} along with a Hubbard correction\cite{Dudarev1998} of $U_\textrm{eff}=8.5$ eV\cite{Chen2017,Kaspar2019a} to describe the Ti $3d$ states. All the calculations were performed using spin-polarization and a $4 \times 4 \times 1$ Monkhorst-Pack\cite{Monkhorst1976} $k$-point mesh to sample the Brillouin zone. The cutoff energy for the projector augmented wave\cite{Blochl1994} pseudo-potential was 500 eV, with convergence criteria of 10$^{-5}$ eV per cell for the energy and 10$^{-3}$ eV \AA$^{-1}$ for the force components. The relaxation procedure for the slab and oxygen vacancies followed the methodology outlined in previous work,\cite{Kaspar2019a} in which the system was first relaxed with PBEsol, and subsequently, a single point calculation of the electronic structure was calculated with PBEsol $+ U$ correction. This was procedure was found to avoid artificial distortions induced by the $U$ correction.

\section{Acknowledgements}

S.R.S. thanks Dr. Colin Ophus for useful discussions. This research was supported by the Laboratory Directed Research and Development (LDRD) Nuclear Processing Science Initiative (NPSI) at Pacific Northwest National Laboratory (PNNL). PNNL is a multiprogram national laboratory operated for the U.S. Department of Energy (DOE) by Battelle Memorial Institute under Contract No. DE-AC05-76RL0-1830. Thin film deposition and characterization was performed using the Environmental Molecular Sciences Laboratory (EMSL), a national scientific user facility sponsored by the Department of Energy's Office of Biological and Environmental Research and located at PNNL. STEM sample preparation and imaging was performed in the Radiological Microscopy Suite (RMS), located in the Radiochemical Processing Laboratory (RPL) at PNNL. Ion irradiations and RBS measurements were carried out at the Texas A\&M University Ion Beam Lab.

\clearpage

\bibliography{references}

\end{document}


\section*{Supplementary Note 1: Rutherford Backscattering Spectrometery (RBS)}

\begin{figure}[h]
\includegraphics[width=0.7\textwidth]{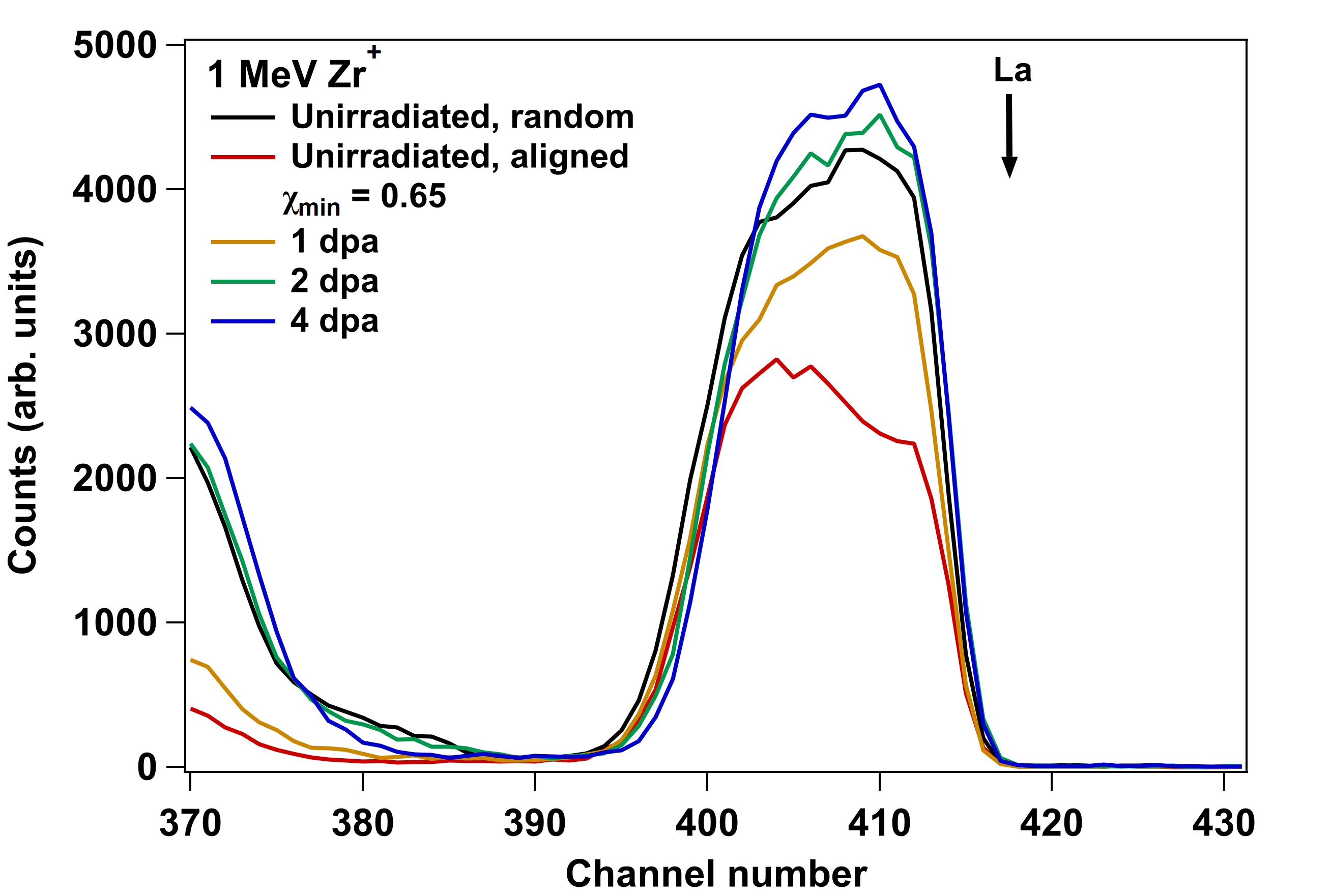}
\caption{Rutherford backscattering spectrometery channeling (RBS/c) measurements showing that most order in the LTO film is lost above a dose of 1 dpa.}
\end{figure}